\documentclass[10pt,leqno]{article}
\usepackage{graphicx}
\usepackage[pagewise, modulo]{lineno}
\usepackage{hyperref}
\usepackage{gensymb}
\usepackage[nointegrals]{wasysym}
\usepackage{mathtools}
\usepackage{setspace}
\usepackage[labelsep=period]{caption}
\usepackage{float}
\floatstyle{plaintop}
\restylefloat{table}

\usepackage{indentfirst,csquotes}

\usepackage{amssymb,amsthm,amsmath}
\usepackage{xcolor,paralist,hyperref,titlesec,fancyhdr,etoolbox}

\hypersetup{ colorlinks=true, linkcolor=black, filecolor=black, urlcolor=black }

\begin{document}
\begin{center}
  \LARGE 
   Detection of $^{135}$Cs \& $^{137}$Cs in environmental samples by AMS \par \bigskip

  \normalsize
  Alexander Wieser\textsuperscript{1,2}, Johannes Lachner\textsuperscript{1,3}, Martin Martschini\textsuperscript{1}, Dorian Zok\textsuperscript{4}, Alfred Priller\textsuperscript{1}, Peter Steier\textsuperscript{1} and Robin Golser\textsuperscript{1} \par \bigskip

  \textsuperscript{1}University of Vienna, Faculty of Physics, Isotope Physics, Währinger Strasse 17, 1090 Vienna, Austria \par
  \textsuperscript{2}University of Vienna, Vienna Doctoral School in Physics, Boltzmanngasse 5, 1090 Vienna, Austria \par
  \textsuperscript{3}Helmholtz-Zentrum Dresden-Rossendorf, Accelerator Mass Spectrometry and Isotope Research, Bautzner Landstraße 400, 01328 Dresden, Germany \par
  \textsuperscript{4}Leibniz University Hannover, Institute for Radioecology and Radiation Protection, Herrenhäuser Strasse 2, 30419 Hannover, Germany \par 
  \bigskip

\end{center}
DOI: https://doi.org/10.1016/j.nimb.2023.02.013 \\ \copyright   This manuscript version is made available under the CC-BY 4.0 license \\ https://creativecommons.org/licenses/by/4.0/

\let\thefootnote\relax
\begin{abstract}
The detection of low abundances of $^{135}$Cs in environmental samples is of significant interest in different fields of environmental sciences, especially in combination with its shorter-lived sister isotope $^{137}$Cs. The method of Ion-Laser InterAction Mass Spectrometry (ILIAMS) for barium separation at the Vienna Environmental Research Accelerator (VERA) was investigated and further improved for low abundance cesium detection. The difluorides BaF$_2^-$ and CsF$_2^-$ differ in their electron detachment energies and make isobar suppression with ILIAMS by more than 7 orders of magnitude possible. By this method, samples with ratios down to the order of $^{135,137}$Cs/$^{133}$Cs $\approx 10^{-11}$ are measurable and the $^{135}$Cs/$^{137}$Cs ratios of first environmental samples were determined by AMS.  
\end{abstract} 

\bigskip \noindent

\section{Introduction}
The long-lived fission product $^{135}$Cs is of interest in the environmental science community \cite{Hain}\cite{Zhu_TIMS} as its measurement adds additional information to the determination of the more commonly used cesium radioisotopes $^{137}$Cs and $^{134}$Cs \cite{CHINO20160822}. Values for the $^{135}$Cs half-life range from 1.3\,Ma to 2.9\,Ma \cite{Cshalflife}\cite{sugarman} and it is a pure beta emitter with a high cumulative fission yield of 6.55\% for $^{235}$U(n$_{\mathrm{th}}$,f). Due to the stability of $^{134}$Xe and the high thermal neutron capture cross section of $^{135}$Xe, the cesium isotopes $^{134}$Cs and $^{135}$Cs are shielded from their beta-decay chain occurring after a nuclear fission process. The isotopic ratio of the shielded cesium isotopes versus the unshielded $^{137}$Cs is dependent on the thermal neutron flux of the fission source and can therefore give information about the anthropogenic source of radioisotopes. However, the applications of the $^{134}$Cs/$^{137}$Cs ratios are limited due to the short half-life of $^{134}$Cs of only 2.07 years. The isotopic ratio including the longer-lived Cs isotopes $^{135}$Cs/$^{137}$Cs can be used in e.g. source term attribution, dating soils and sediments \cite{LEE19933493} and environmental transport models \cite{YANG2016177}. The fission yield ratio of $^{135}$Cs/$^{137}$Cs is 1.06, making it impossible to measure $^{135}$Cs in environmental samples via radiometric methods due to the overwhelming beta activity of $^{137}$Cs or $^{134}$Cs. In the past years, several mass spectrometric methods have shown to be very successful in determining the $^{135}$Cs/$^{137}$Cs ratios in contaminated areas down to abundance sensitivities of $^{135}$Cs/$^{133}$Cs = $10^{-10}$. However, as we show in the following, AMS has the potential of reaching isotopic ratios 3-4 orders of magnitude lower. This is needed for analysis of general environmental samples \cite{ZHAO201696} by suppressing both the peak-tailing from $^{133}$Cs with various momentum and energy filters and the interfering molecules with mass 135 and 137 with a tandem accelerator. First AMS measurements of $^{135}$Cs were performed by Zhao et al. and Macdonald et al. using difluoride molecules \cite{ZHAO201696}\cite{MACDONALD2015554}. Here, we describe the suppression of the atomic isobars $^{135,137}$Ba with the Ion-Laser InterAction Mass Spectrometry (ILIAMS) setup at VERA \cite{ILIAMS} also using difluoride molecules (CsF$_2^-$, BaF$_2^-$) and finally, results of measurements of first environmental samples showing the $^{135}$Cs/$^{137}$Cs signature from the Chernobyl and Fukushima nuclear accidents by AMS. The values obtained by VERA are compared to values already measured by ICP-QQQ-MS at the University of Hannover \cite{zok}. 

\section{Materials \& Methods}
\subsection{Internal Reference Material}
There is neither a certified $^{135}$Cs/$^{133}$Cs nor a $^{135}$Cs/$^{137}$Cs standard material. There exists, however, consensus values on some IAEA and NIST reference materials (see e. g. \cite{BUETAL}\cite{YANG2016177}\cite{SNOW201517}\cite{SNOW2016_156}). At VERA, we use an in-house reference material solution with an a priori unknwon $^{135}$Cs/$^{137}$Cs ratio but a measurable $^{137}$Cs activity of 1\,Bq/ml. This reference material was measured over several years by AMS at VERA to have $^{135}$Cs/$^{137}$Cs = 2.82 $\pm$ 0.12 (decay corrected to March 1$^{\mathrm{st}}$, 2021) assuming equal detection efficiency for $^{137}$Cs and $^{135}$Cs. This reference material is mixed with a certain amount of Cs carrier (Cs$_2$SO$_4$ in Milli-Q water) to obtain the desired $^{137}$Cs/$^{133}$Cs ratios and further mixed with PbF$_2$ and Cu powder and, once dried, pressed into Cu sample holders with a Cu pin. These reference samples then are used to determine the normalization factor for the AMS measurement. The stable Cs content of the initial reference solution was determined by ICP-MS to be in the ng/ml range and is therefore negligible for AMS measurements, as typically $\approx$ 1\,ml is mixed with $\approx$ 10\,mg of Cs carrier.

\subsection{Environmental samples}
Two samples, moss collected in Fukushima and a catfish caught in the Chernobyl exclusion zone, were chemically prepared and measured by ICP-QQQ-MS at Hannover \cite{zok} and further prepared as AMS targets in Vienna, by adding 1\,mg of Cs-carrier (Cs$_2$SO$_4$ in Milli-Q water), 3\,mg of PbF$_2$ and 1\,mg of Cu-powder to the sample solutions. The samples were then dried for several hours in teflon beakers and pressed into Cu sample holders, just as for the reference material mentioned above. The $^{137}$Cs/$^{133}$Cs for the moss AMS samples was $1.3\cdot10^{-10}$ and for the catfish $6.0\cdot10^{-10}$. These ratios were determined by gamma spectrometry \cite{zok} of the sample and weighing the admixed carrier solution. 

\subsection{Sputtering Cs samples} 
A major challenge for Cs measurements with AMS is the sputtering process, which is done with Cs cations in a Cs sputter ion source. There are different approaches in solving this problem, by spiking $^{134}$Cs into the sample \cite{MACDONALD2015554}, using no sputter agent at all \cite{ZHAO201696} or using rubidium (Rb) sputtering. The latter one was applied at VERA for some previous experiments \cite{LACHNER2015440} and was also used for all samples in this study. Prior to each Cs beamtime, the ion source is cleaned thoroughly to avoid any Cs contamination from previous beamtimes, in which Cs sputtering was applied. Further, ion source parts like the Cs/Rb reservoir and used materials such as the Cs carrier, PbF$_2$ and Cu powder were all checked for $^{137}$Cs activity by gamma spectrometry. As we describe below, the $^{137}$Cs AMS detection limit at VERA is in the mBq range using mg amounts of sample material. If this background level was due to contaminations of the above mentioned materials, one should see activities in the Bq range for gram amounts of these materials. Since this was clearly not the case, the $^{137}$Cs counts on the blank material can not be attributed to any intrinsic $^{137}$Cs in the used materials.

\subsection{ILIAMS}
Ba and Cs are not separable at VERA in a gas-filled ionization chamber, even at energies of 26.2 MeV \cite{LACHNER2015440}. Therefore, the low-energy isobar suppression method ILIAMS is used to detect Cs. The ILIAMS setup is centered around a He buffer-gas filled radiofrequency quadrupole, in which the ion beam is electrostatically decelerated to some tens of eV and further cooled to near-thermal energies by the buffer-gas \cite{FORSTNER2015217}\cite{MARTSCHINI20179}. In this quadrupole trap the ion beam is collinearly overlapped with a laser beam. The principle of this method is to use a laser with a photon energy, that exceeds the electron detachment energy of the unwanted isobar (Ba), but is still lower than the electron detachment of the element of interest, in our case Cs. The vertical detachment energy, which is the minimum energy needed to eject an electron from the anion in its ground state without changing the internuclear distance, of Cs$^-$ is 0.47\,eV \cite{Cs_EA}, while that of Ba$^-$ is only 0.14\,eV \cite{Ba_EA}. A laser with photon energy in between those two energies would suppress the Ba according to
\begin{equation}
    N_s=N_0 \cdot e^{-\Phi \sigma(\lambda) t}
\end{equation}
where N$_0$ is the number of injected ions, $\Phi$ is the photon flux, $\sigma(\lambda)$ is the photodetachment cross section dependent on the wavelength (typically Mb), t is the interaction time of laser and ion beam (typically ms) and N$_\mathrm{s}$ is the number of ions surviving the interaction. For photon energies below the detachment energy, the cross section becomes zero. Despite the favorable detachment energies, Cs$^{–}$ is not a good choice for AMS measurements, since it gives only poor ion beams on the order of 1-5 nA \cite{LACHNER2015440}. Hence, a molecular system needs to be found, for which the Cs output is higher, but the detachment energies are still suitable for ILIAMS. As shown by Zhao et al. \cite{ZHAO201696}, CsF$_{2}^{–}$ as a superhalogenic molecule readily forms anions. Further, it makes the suppression of BaF$_{2}^-$ with a green laser of 2.33 eV (532 nm) photon energy possible \cite{5yearsofiliams}. CsF$_{2}^{–}$ is produced in the ion source from cathodes containing a mixture of Cs$_2$SO$_4$ with PbF$_2$ and gives ion currents of 50-100 nA with Rb sputtering. The calculated vertical electron detachment energies for CsF$_{2}^-$ and for BaF$_{2}^-$ are 3.88\,eV and 0.61\,eV, respectively \cite{Armin}. This is in accordance with our first experiments: With our available lasers at wavelengths 532\,nm and 355\,nm we could confirm that the vertical detachment energy for CsF$_{2}^-$ is higher than 3.49\,eV and that of BaF$_{2}^-$ is lower than 2.33\,eV.

\subsection{Accelerator Mass Spectrometry for Cs at VERA}

\begin{figure}[h!]
    \centering
    \includegraphics[width=0.5\textwidth]{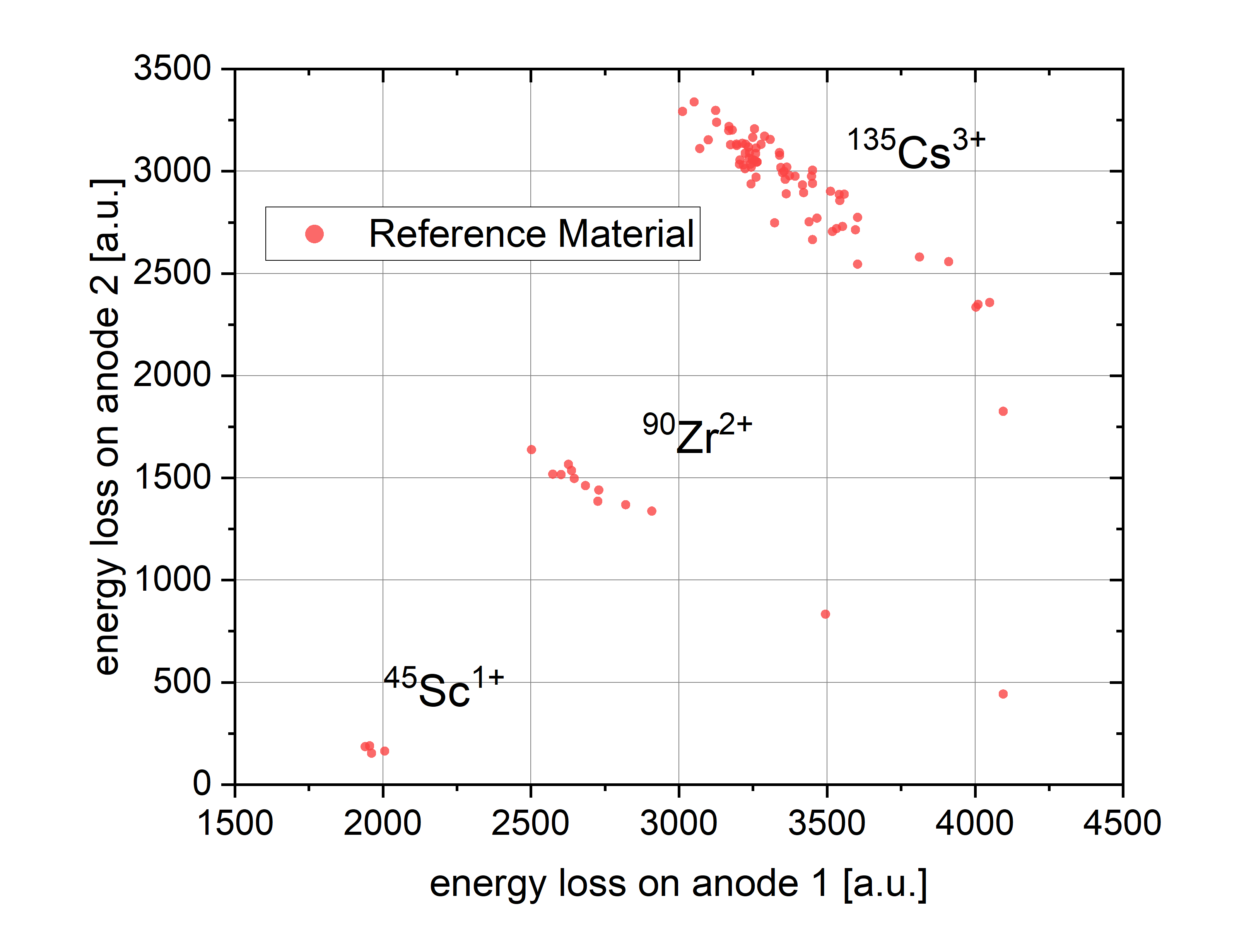}
    \caption{$^{135}$Cs$^{3+}$ can be easily separated from the m/q interferences $^{45}$Sc$^{1+}$ and $^{90}$Zr$^{2+}$ by their energy loss behavior in a gas-filled ionization chamber. To be detectable on the high-energy side, Zr and Sc need to form molecules with mass 173 in the ion source, which have a high detachment energy and can survive the laser photodetachment with the green laser.  }
    \label{gic}
\end{figure}

For Cs measurements at VERA CsF$_2^-$ anions are extracted with an ion source voltage of 30\,kV. Then, they are mass-analyzed with a 90$\degree$ bending magnet before they are decelerated again and injected into the ion cooler. $^{133}$CsF$_2^-$ currents on the order of 10-100\,nA are produced, so no attenuation of the ion beam is necessary. After passing the ion cooler, the surviving anions are reaccelerated to 30\,keV and injected into the AMS system. With a terminal voltage of 2.6\,MV and helium stripper gas, the 3+ charge state is selected and has an accelerator transmission of 23\%. The $^{133}$Cs$^{3+}$ current on the order of 10\,nA is measured after the high-energy magnet in an offset Faraday cup. The isotopes with mass 135, 136 and 137 are sequentially measured in a gas-filled ionization chamber (GIC) after they pass the high-energy ESA. For mass 135 in the 3+ charge state, m/q interferences of $^{90}$Zr$^{2+}$ and $^{45}$Sc$^{1+}$ appear but can be easily separated from Cs/Ba in the GIC (see fig. \ref{gic}) . Mass 136 is measured for monitoring the barium suppression as only stable barium and cerium should occur. CeF$_2^-$ does form anions and is indeed present in either the Cs carrier or the Ba carrier, but the detachment energy of CeF$_2^-$ is below 2.33\,eV, so it is also completely suppressed by ILIAMS. The Cs measurement is done in a slow-sequencing mode, where the stable isotope is injected into the cooler for about 10 seconds, followed by the three rare isotope masses for about 100-200 seconds each. This sequence is repeated three times before switching to the next sample. For an overview of the VERA setup see e.g. \cite{STEIER200467}. 

\section{Results and Discussion}
\subsection{Barium suppression with ILIAMS}
\begin{figure}[h!]
    \centering
    \includegraphics[width=0.5\textwidth]{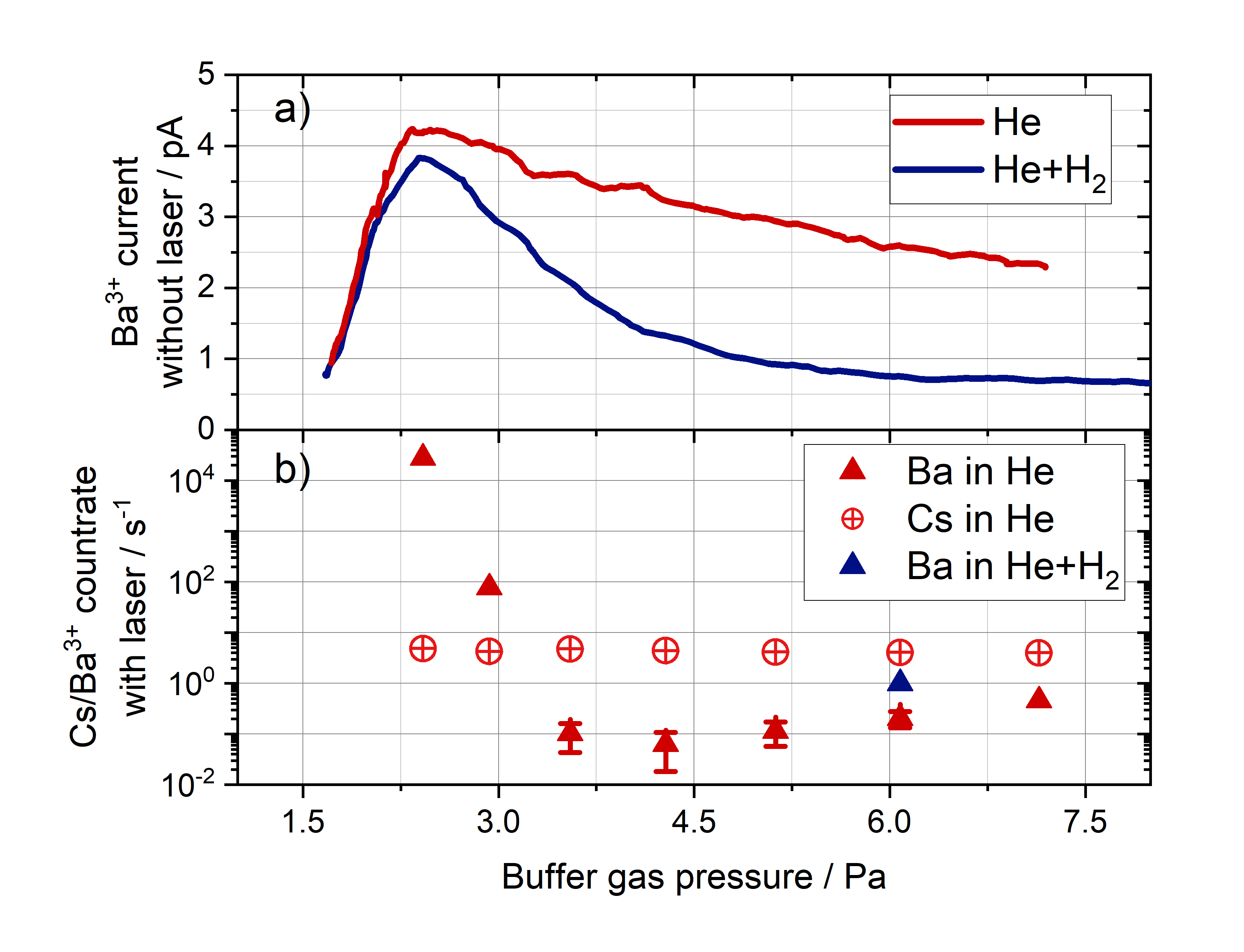}
    \caption{$^{137}$Ba$^{3+}$ current (without the laser in panel a)) measured in the Faraday cup in front of the ionization chamber used for Cs measurements. Barium is clearly more suppressed by buffer gas containing 10\% of H$_2$. However, this effect does not hold up when adding the laser to the system (see panel b)). For the overall suppression, we do not see any improvement by adding H$_2$ to the buffer gas. The Cs countrate on the in-house reference material is not affected by the buffer gas pressure. Where not depicted, the error bars, stemming from counting statistics, are smaller than the symbol size. }
    \label{heh2_barium}
\end{figure}

The primary goal of the ion cooler is to maximize the interaction time between the laser beam and the ion beam but it can also act or be used as a gas reaction cell. This opens up several ways to suppress isobars with the ILIAMS setup e.g. via molecular reactions such as attaching an oxygen-, hydrogen-, or fluorine-atom or dissociating in the ion cooler. Data on the amount of BaF$_2^-$ transmitted from a barium sample through the ILIAMS cooler, with and without laser, is shown in fig. \ref{heh2_barium} for two different buffer gases, pure He and He+H$_2$. The overall suppression of BaF$_{2}^-$ was determined to be $2 \cdot 10^7$ on a BaF$_{2}$+PbF$_2$ target, where we could detect pA of $^{137}$Ba$^{3+}$ without the laser and approximately 0.1 counts per second with laser and with buffer gas pressures above 4.5\,Pa. The pressure values given here are calculated for the center of the ion cooler. The average pressure in the cooler tube is approximately a factor two lower than that, as the He density declines step wise towards the cooler tube openings \cite{Gaggl}. Below 4.5\,Pa, the countrate is strongly increasing by several orders of magnitude, due to insufficient interaction time of ion and laser beam. Below 2.0\,Pa, the ions are not well enough thermalized to reliably pass the 3\,mm exit aperature of the cooler and beam losses increase. These ion optical losses will occur both for Cs and Ba and hence do not provide any isobar separation. In order to achieve optimal Ba suppression the focusing and positioning of the laser beam is critical and sensitive. This tuning of the laser beam is typically performed reaching very small countrates of $\approx$ 0.01\,–\,0.1 s$^{-1}$ on barium spiked samples on mass 136. Due to these low count rates, the quality of the laser tuning and thus the overall suppression factor may vary by up to one order of magnitude between different beam times. Together, with a suppression of CsF$_{2}^-$/BaF$_{2}^-$ $\approx$ 20 in the ion source \cite{LACHNER2015440}, we reach an overall suppression of $4 \cdot 10^8$ of barium at VERA, while reaching detection efficiencies for Cs on the order of 0.1$\permil$ (30\% cooler transmission, 23\% accelerator transmission, 1$\permil$ ionization efficiency). \\
\\
Towards higher pressures the dominant suppression channel for barium seems to be the neutralization by laser light (BaF$_{2}^{-}$ + $\gamma$ $\rightarrow$ BaF$_{2}$ + e$^{-}$) as no molecular changes such as attaching an oxygen-, hydrogen or fluorine-atom or dissociating in the ion cooler could be observed. A marginally stronger suppression of barium by a helium-hydrogen mixture as buffer gas, compared to pure helium (see fig. \ref{heh2_barium} panel a)) was found which will be investigated further in the coming beamtimes, whether a molecular reaction from BaF$_{2}^-$ to BaF$_{2}$H$^-$ is the main cause for the decreasing ion current. However, in general, the suppression by the gas only and the combined suppression by the gas + laser may show completely different trends, especially when a change of the molecular system of the isobar in the ion cooler happens. Since some of the molecules formed via reactions in the cooler will have a high electron detachment energy, they are not affected by the laser and may pass the cooler unaffected. In order to interfere in the measurement, these "Trojan horse" molecules need to do a back reaction somewhere near the cooler exit, which obviously occurs with low but not negligible probability. In fact, the isobaric suppression with additional laser photodetachment turned out to be of equal magnitude for the helium-hydrogen mixtured buffer gas. For BaF$_2^{-}$, we believe these "Trojan horses" to be either BaF$_{2}$H$^-$ or BaF$_{3}^-$, where the latter was verified to have a detachment energy above 3.49\,eV, by injecting BaF$_{3}^-$ into the cooler. The additional fluorine atom could be provided by a collisional breakup of fluorine-containing molecules including CsF$_{2}^-$ + He $\rightarrow$ CsF + F$^-$ + He. The F-breakup of CsF$_{2}^-$ was found to happen by injecting CsF$_{2}^-$ into the cooler, setting the low-energy magnet to mass 19 and searching for F$^{1+}$ on the high-energy side of the AMS system. Still, the cooler transmission for CsF$_{2}^-$ is about 30\% and widely insensitive to the buffer gas pressure, which is similar to other fluoride molecules (CaF$_3^-$, SrF$_3^-$, HfF$_5^-$) at VERA. A similar reaction for the isobar, BaF$_{2}^-$ + He $\rightarrow$ BaF + F$^-$ + He, was not detected. 

\subsection{Reproducibility and cross contamination}
\begin{figure}[h!]
    \centering
    \includegraphics[width=0.5\textwidth]{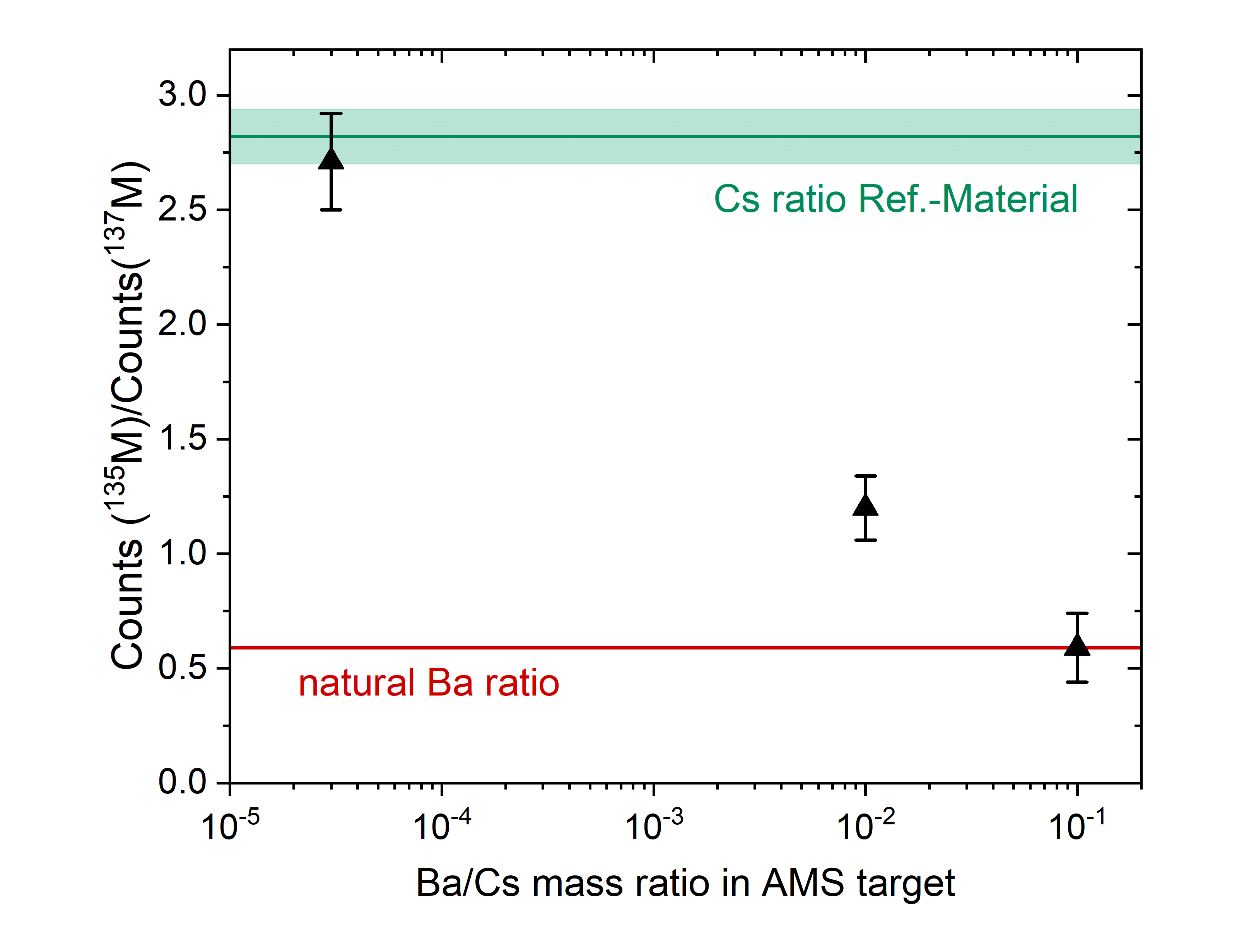}
    \caption{The ratio of counts on mass 135 to mass 137 depends on the Ba content of the AMS target. These values are obtained by long-term averages over many beamtimes. }
    \label{bariumconc}
\end{figure}

Besides determining the normalization factor for the AMS measurement, the internal reference material can also be used to differentiate between counts stemming from cross contamination from the reference material versus insufficient barium suppression on the used blanks. Since the natural barium ratio is $^{135}$Ba/$^{137}$Ba = 0.59, these two effects are quite easily distinguishable (see fig. \ref{bariumconc}). Ba spiked samples (Cs$_2$SO$_4$ + BaF$_2$ + PbF$_2$) containing 10\% or more Ba by mass have a low 135/137 ratio, meaning the barium suppression is not complete. For the Cs carrier, containing nominally $< 0.003\%$ barium and no $^{137}$Cs, this ratio is dominated by cross contamination from the reference material. For samples containing 1\% barium by mass, these two effects are mixing. We expect to suppress the barium in environmental samples with barium mass contents of M$_{\mathrm{Ba}}$/M$_{\mathrm{Cs}}$ $< 1\permil$ after chemical treatment. \\

\begin{figure}[h!]
    \centering
    \includegraphics[width=0.5\textwidth]{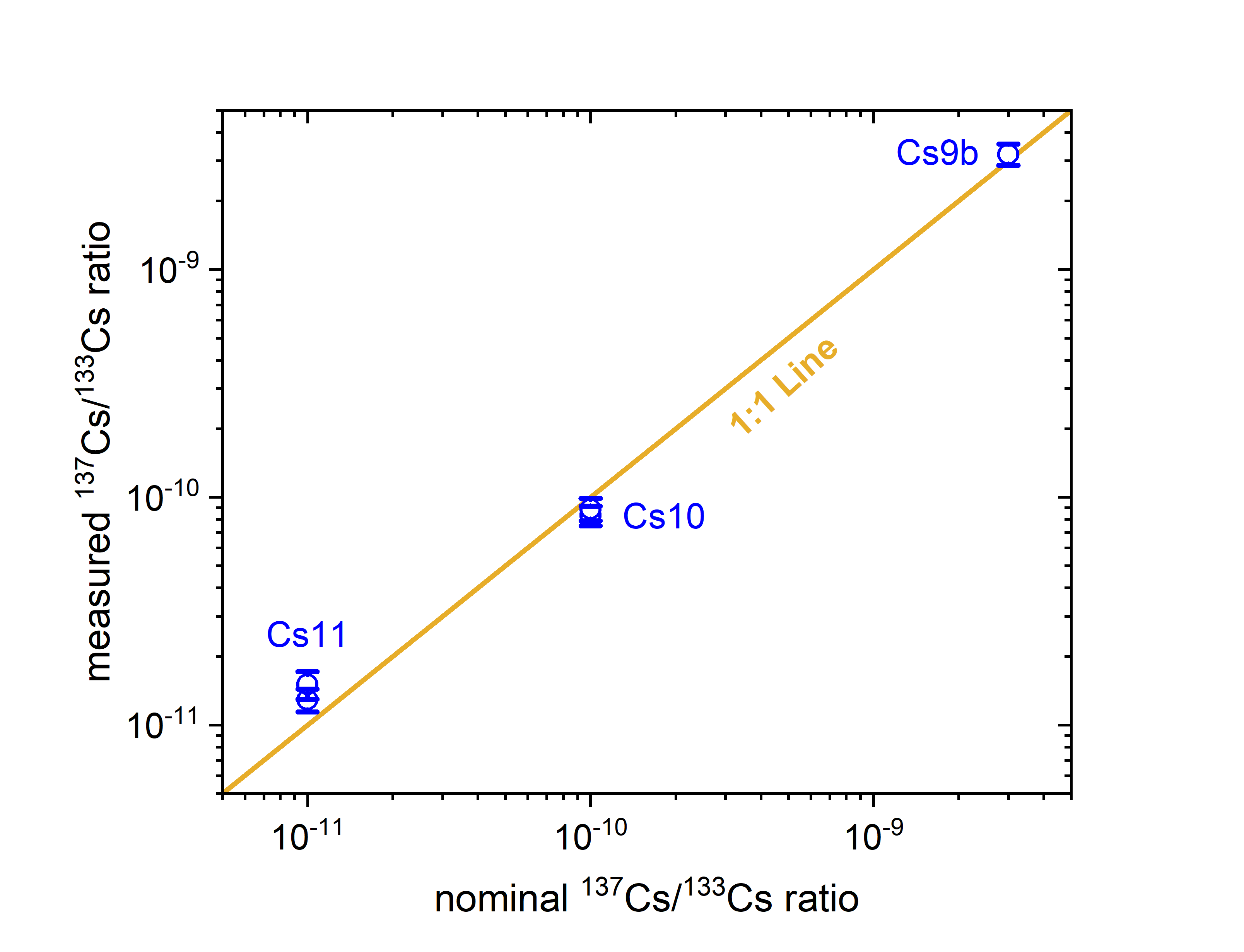}
   \caption{The displayed isotopic ratios are measured on the internal reference material, which was admixed with different amounts of stable Cs carrier. The ratios are normalized to Cs9b ($^{137}$Cs/$^{133}$Cs=3$\cdot10^{-9}$) and Cs10 ($^{137}$Cs/$^{133}$Cs=1$\cdot10^{-10}$). }
    \label{linearity}
\end{figure}

The linearity of a dilution series of the reference material was tested in a single beamtime with $^{137}$Cs/$^{133}$Cs ratios spanning over more than two orders of magnitude (see fig. \ref{linearity}). The elevated isotopic ratios for lower samples can be explained by cross contamination on the order of 5\%, which in this test already affects samples with nominal ratios of $^{137}$Cs/$^{133}$Cs = $1\cdot10^{-11}$. The severity of the cross contamination is about a factor of six higher than for Cl$^{-}$ at VERA \cite{PAVETICH201422}, where an average contamination of 0.8\% was measured. This relatively high sensibility of Cs regarding cross contamination prevents reproducible Cs detection for samples spanning over more than two orders of magnitude in their radiocesium abundances. The choice of reference material for real samples is therefore crucial. For environmental samples containing more than 100\,mBq of $^{137}$Cs, this selection can be done based on gamma spectrometry. The cross contamination was found to be strongly reduced by adding copper powder to the sample material, potentially due to improved thermal and electrical conductivity of the AMS target.

\subsection{Blank values and detection efficiency}
\begin{figure}[h]
    \centering
    \includegraphics[width=0.5\textwidth]{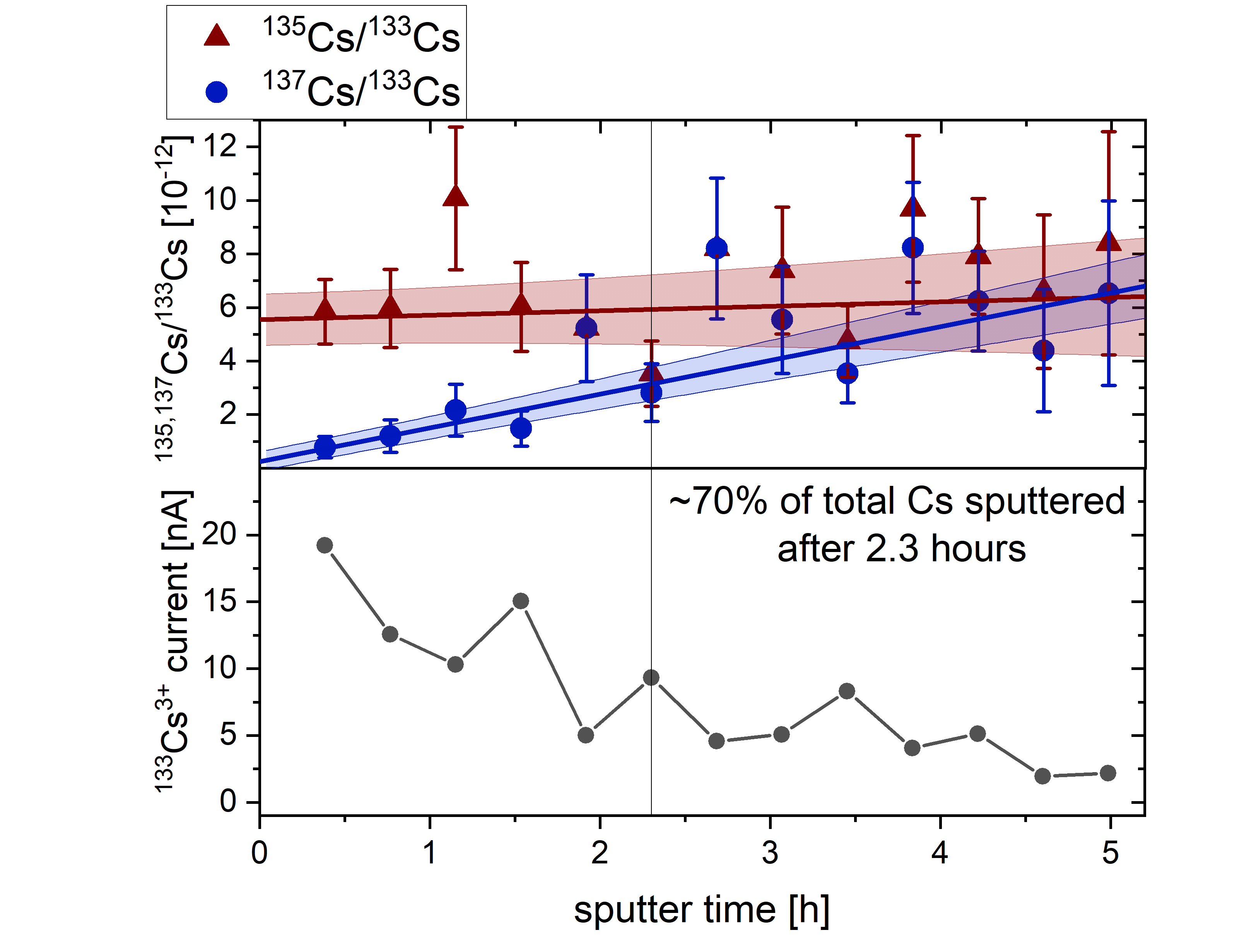}
    \caption{Blank values on both isotopes of one blank sample in consecutive runs. The blank sample and a reference material with $^{137}$Cs/$^{133}$Cs = $1\cdot10^{-10}$ were alternately measured in this test. The $^{135}$Cs/$^{137}$Cs ratio starts very high, which we interpret as intrinsic $^{135}$Cs input from the Cs carrier, the PbF$_2$, Cs/Rb oven or the copper cathodes. Intrinsic $^{137}$Cs in all used materials was ruled out by gamma spectrometry. The grey line indicates that after 2.3 hours of sputtering roughly 70\% of the sample is used up. }
    \label{blanksovertime}
\end{figure}

With the present setup we reach blank values for the Cs$_2$SO$_4$-carrier (Alfa Aesar, 99.997\% (metals basis), LOT No. S94194) of $^{135}$Cs/$^{133}$Cs = $6\cdot10^{-12}$ and $^{137}$Cs/$^{133}$Cs = $3\cdot10^{-12}$ with $^{135,137}$Cs counts stemming from cross contamination from samples in the $10^{-10}$ range dominating clearly on the non-barium spiked targets. This contamination effect also leads to increasing blank values over measurement time (see fig. \ref{blanksovertime}). For another carrier (LGC ICP Cs Standard 1000\,$\mu$g/ml, Cs$_2$CO$_3$ in 5\% HNO$_3$, LOT No. 1191949-11) a dedicated blank test was done by sputtering the blanks over several hours. Two samples (LGCa and LGCb) were measured for their $^{135}$Cs/$^{137}$Cs ratio, followed by one Cs10 ($^{137}$Cs/$^{133}$Cs = $1\cdot10^{-10}$) sample for checking the validity of the setup and introducing as little cross contamination as possible. This LGCa–LGCb–Cs10 sequence was repeated six times. In this test, blank ratios of $^{137}$Cs/$^{133}$Cs = $(6.0\pm2.7)\cdot10^{-13}$ (5 counts in 2700\,s) and $^{135}$Cs/$^{133}$Cs = $(4.77\pm0.76)\cdot10^{-12}$ (40 counts in 2700\,s) were obtained. The high $^{135}$Cs/$^{137}$Cs ratio of $8.0\pm3.8$ leads to the suspicion that we suffer from constant $^{135}$Cs input from one of the used materials or the ion source setup. For 1\,mg of Cs, the blank value for $^{137}$Cs translates to a detection limit of 4-5\,mBq or $6\cdot 10^6$ atoms (Blank value + 3 times the uncertainty) per AMS target, which is about one order of magnitude higher than detection limits in underground laboratories (see e.g. \cite{gamma}). For $^{135}$Cs, this abundance sensitivity is 1-2 orders of magnitude below other mass spectrometric methods such as TIMS and ICP-MS \cite{Zhu_TIMS}\cite{zhu_icp}. \\
\\
To detect 1000 counts of $^{135,137}$Cs, we need the sample to contain $\approx$ $3\cdot10^7$ atoms of each isotope, with ideal ion source output. Despite the detection efficiency of $^{133}$Cs being 0.1$\permil$, the detection efficiency for the radioisotopes is a factor three lower. The reason for this major loss of the radioisotopes compared to the stable Cs is not fully understood yet but may be related to a difference in residence time in the ion cooler for intense and rare isotope beams \cite{MARTSCHINI20179}. 
\subsection{Environmental samples measured by ILIAMS and AMS}

First environmental samples measured at VERA were compared with values obtained by ICP-QQQ-MS at the University of Hannover (see tab. \ref{environmental_samples}) \cite{zok}. The samples were measured together with our internal Cs10 ($^{137}$Cs/$^{133}$Cs = $1\cdot10^{-10}$) reference material and blanks, containing only the Cs carrier mixed with PbF$_2$ and Cu powder.\\
The $^{135}$Cs/$^{137}$Cs ratios obtained in Hannover are mean values over several measurements. In a first approach, the AMS (raw data) values are not normalized to any reference material, as the $^{137}$Cs/$^{133}$Cs normalization factor should be the same for $^{135}$Cs/$^{133}$Cs and thus cancels out for calculating the $^{135}$Cs/$^{137}$Cs ratio. The $^{135}$Cs/$^{137}$Cs of the in-house reference material in this beamtime was 2.97 $\pm$ 0.19, which is in 1\,$\sigma$ agreement with our long-term average. Alternatively, if one corrects the values obtained for the environmental samples with the long-term reference value, one gets for the Moss $0.433 \pm 0.048$ and for the Catfish $0.67 \pm 0.13$, which is still in agreement with the values measured by ICP-QQQ-MS. This also indicates that the counting of $^{137}$Cs and $^{135}$Cs in the AMS and ILIAMS system does not suffer from any major systematic uncertainties. 

\section{Conclusion}
The ILIAMS setup at VERA is able to detect $^{135}$Cs and $^{137}$Cs by suppressing barium via laser-photodetachment. Blank ratios on the order of $10^{-12}$ are achieved with extracting CsF$_2^{-}$ anions by sputtering with rubidium cations. Cross contamination in the ion source and the generally low ion source output of only 50-100\,nA of CsF$_2^{-}$ currently prevent us from reaching lower ratios. Thanks to the excellent Ba-suppression by ILIAMS of more than 7 orders of magnitude, the isobaric interferences are significantly lower than $10^{-12}$. \\
\\
Further, first environmental samples were measured by AMS at VERA showing the $^{135}$Cs/$^{137}$Cs signature of the nuclear accidents in Fukushima and Chernobyl. We could reproduce the $^{135}$Cs/$^{137}$Cs values obtained by ICP-QQQ-MS within 1\,$\sigma$ uncertainty. AMS is capable of measuring low isotopic ratios in samples, where high amounts of stable Cs are present, as we do not suffer from any peak tailing of $^{133}$Cs to the other masses. Future goals are to lower the blank by improving not only chemical preparation of the AMS targets but also by understanding the cross contamination problem. The barium suppression by laser photodetachment is sufficiently high for samples containing 1$\permil$ of barium when measuring samples with $\approx$\,Bq of $^{137}$Cs but will need to be improved when measuring general environmental samples with isotopic ratios in the $^{135,137}$Cs/$^{133}$Cs\,$\approx 10^{-14}$ range.  

\section{Acknowledgements}
We thank Armin Shayeghi (University of Vienna) for calculating the vertical detachment energies for the difluoride molecules and Sabrina Beutner (Helmholtz-Zentrum Dresden-Rossendorf) for determining the stable Cs concentration in the $^{137}$Cs reference solution. \\
Part of this work was funded by the RADIATE project under the Grant Agreement 824096 from the EU Research and Innovation program HORIZON 2020. The authors acknowledge financial support of the ILIAMS research activities by “Investitionsprojekte” of the University of Vienna.

\bibliographystyle{unsrt}
\bibliography{mybibfile}

\section{Tables}

\begin{table}[h]
    \centering
    \begin{tabular}{c|c|c}
         & Moss (Fukushima) & Catfish (Chernobyl) \\ \hline
    ICP-QQQ-MS & 0.445 $\pm$ 0.005 & 0.625 $\pm$ 0.005 \\
    AMS (raw data) & 0.456 $\pm$ 0.037 & 0.70 $\pm$ 0.12 \\
    AMS (normalized) & 0.433 $\pm$ 0.048 & 0.67 $\pm$ 0.13  
    \end{tabular}
    \caption{$^{135}$Cs/$^{137}$Cs ratios measured by ICP-QQQ-MS at the IRS Hannover \cite{zok} and by AMS at VERA decay corrected to March 1$^{\mathrm{st}}$, 2021. For both samples, two AMS targets were produced and averaged. The higher uncertainty for the Catfish stems from lower ion source output.}
    \label{environmental_samples}
\end{table}

\newpage
\section{Figures}

\end{document}